\documentclass[aps,prd,showpacs,amssymb,superscriptaddress, notitlepage,floats,floatfix,nofootinbib, twocolumn]{revtex4-2}

\usepackage{booktabs}
\usepackage{graphicx}
\usepackage{subfig}
\usepackage{amssymb}
\usepackage{amsmath}
\usepackage{amsfonts}
\usepackage{tensor}
\usepackage{bm}
\usepackage{mathrsfs}
\usepackage[colorlinks=true]{hyperref}
\usepackage[all]{hypcap}
\usepackage[utf8]{inputenc}
\usepackage[normalem]{ulem}
\usepackage{url}

\usepackage{float}

\usepackage{xcolor}
\usepackage{multirow}
\usepackage{palatino}
\usepackage{float}
\usepackage{siunitx}

\usepackage{journals}
\usepackage{soul}
\usepackage{calrsfs}

\begin{document}
\title{The gravitational Higgs mechanism and resulting smoking gun effects}
\author{Verena Krall}
\affiliation{Theoretical Astrophysics, IAAT, University of T\"ubingen, Germany}
\affiliation{Chair for Network Dynamics, Institute for Theoretical Physics and Center for Advancing Electronics Dresden (cfaed), Technical University of Dresden, Germany}
\email{verena.krall@tu-dresden.de}
\author{Andrew Coates}
\affiliation{Theoretical Astrophysics, IAAT, University of T\"ubingen, Germany}
\affiliation{Department of Physics, Ko\c{c} University,
Rumelifeneri Yolu, 34450 Sariyer, Istanbul, Turkey}
\email{acoates@ku.edu.tr}
\author{Kostas D. Kokkotas}
\affiliation{Theoretical Astrophysics, IAAT, University of T\"ubingen, Germany}
\date{\today}
%
\begin{abstract}
Recently, a toy model was introduced to demonstrate that screening mechanisms in alternative theories of gravitation can hide additional effects. In this model a scalar field is charged under a \(U(1)\) symmetry. In sufficiently compact objects the scalar field spontaneously grows, i.e. the object scalarizes, spontaneously breaking the \(U(1)\) symmetry. Exactly as in the \(U(1)\) Higgs mechanism this leads to the emergence of a mass for the gauge field. 

The aim of this paper is to provide an example of the physical consequences if we consider this toy model as a prototype of Weak Equivalence Principle (WEP) violations.
We model neutron stars with a dipolar magnetic field to compare the magnetic field behaviour of stars in Einstein-Maxwell theory on the one hand and in scalar-tensor theory with the, so-called, gravitational Higgs mechanism on the other hand.
\end{abstract}
\maketitle
\section{Introduction}\label{Introduction}
General Relativity (GR) has been tested many times during the hundred years of its existence \cite{Will}, including, since September 2015, via direct gravitational wave observations. All of the observations made with the ground-based laser interferometers LIGO and Virgo have so far been in accordance with the predictions of GR \citep{GW150914, GWBBH, GW170814, GW170817, GW170104}\,. 

Nevertheless, as Karl Popper has most famously stated in \cite{Logik}\,, in empirical sciences it is never possible to verify a theory. On the contrary, what makes a theory worth studying is that it leads to hypotheses which, by experiments, can be tested and falsified. 

In other words, the number of tests passed by GR neither makes it the true theory of gravitation nor removes the necessity to perform more experiments and probe its hypotheses. 

Furthermore, there are questions which have not yet been answered in a satisfactory way by GR: What exactly is dark matter? Why does the cosmological constant have its observed value? And, from a more holistic viewpoint on physics, how can we find a theory of quantum gravity which describes phenomena on all scales at all times?

In order to judge experimental outcomes and to design new experiments, it is important to know which deviations to expect. Stated differently, putting a theory to the test becomes a lot easier if one has an alternative theory at hand which makes some assumptions itself.

One of the simplest alternatives is scalar tensor theory (STT), the first example of which was introduced in 1961 by C. Brans and R. H. Dicke \cite{BD:61}. By introducing an additional scalar field into the Einstein-Hilbert action, one can make alternate predictions with a minimum of additional ingredients. It further offers mechanisms to avoid deviations from GR in the Solar system while still predicting large differences for other regimes where less constraints exist \citep{SpontScal1,SpontScal2,Khoury:2013yya,Hinterbichler:2010es}. These are ``screening mechanisms''. 

Introduced in \citep{GravHiggs1,GravHiggs2}\,, STT with a gravitational Higgs mechanism provides a new perspective on this prominent alternative theory of gravitation. It shows that a deviation from GR need not only lead to changes in the laws of gravity, but it might also trigger other effects, for example, in particle physics. For demonstrative purposes the scalar field was coupled to a \(U(1)\) field in \cite{GravHiggs1}, which then through the Higgs mechanism makes the gauge field massive.

The important fact here is that tests of the Weak Equivalence Principle (WEP) in the solar system do not automatically imply its validity in the strong gravity regime. What makes this interesting is that there are theoretical reasons to believe the WEP should not be generally valid, either through effective field theory arguments - the WEP is not a symmetry so any couplings it forbids should be generated quantum mechanically - or by noting its absence (or even complete incoherence) in most quantum gravity candidates. There is, however, a large issue if one decides to give up the WEP. The number of allowed couplings would be enormous. Additionally they would have to generically be introduced at the level of the standard model of particle physics and getting from there to a neutron star model would be an even more difficult task than it already is (see e.g. \cite{Baym:2017whm}).

The model introduced in \cite{GravHiggs1} has several short-comings if one wishes to interpret it as a serious theory of nature, but it is more than sufficient as a toy model to demonstrate the principle that the WEP in the strong and weak gravity regimes can be decoupled. Our goal here is to use this model to understand the size of some observable differences in neutron stars for even very mild WEP violations, to motivate study of the harder problem described above. To this end, we first give a brief overview of the theoretical background needed in section \ref{Theory}. 

In section \ref{NSModels} we then show that a non-vanishing photon mass in STT with the gravitational Higgs mechanism leads to changes in the magnetic fields of scalarized neutron stars. Finally, we will summarize and discuss our results in section \ref{Discussion}\,.

Throughout the article, unless stated otherwise,  we use units defined by $c=G=1$ and $\mu_0=4\pi$\,. 

\section{Theoretical background}\label{Theory}

\subsection{Scalar-tensor theory}
As the name suggests, STT is a theory of gravity with a metric and a scalar field.
In the standard formulation of STT, the scalar field $\varphi$ couples non-minimally to the Ricci scalar \citep{SotGrav, Berti, STTofGrav}:
\begin{align}
S_\text{J}\,=\,&\frac{1}{16\pi G}\int \mathrm{d}^4 x\sqrt{-g}\left(\varphi R-\frac{\omega(\varphi)}{\varphi}\nabla^\mu\varphi\nabla_\mu\varphi-V(\varphi)\right)\nonumber\\
&+S_\text{m}\lbrack g_{\mu\nu},\psi\rbrack\, . \label{STT_action}
\end{align}
Here, $\omega(\varphi)$ determines the coupling of the scalar field to curvature and $V(\varphi)$ is the potential of the scalar field. After a conformal transformation to the so-called Einstein frame $(\tilde{g}_{\mu\nu}, \phi)$, with $g_{\mu\nu}\rightarrow\tilde{g}_{\mu\nu}=A^{-2}(\phi) g_{\mu\nu}$\,, and where $\phi$ is defined implicitly through $A(\phi)=\varphi^{-1/2}$\,, the action is given by: 
\begin{align}
S_\text{E}\,=\, &\frac{1}{16\pi G}\int \mathrm{d}^4 x\sqrt{-\tilde{g}}\left(\tilde{R} -2\tilde{g}^{\mu\nu}\nabla_\mu\phi\nabla_\nu\phi-U(\phi)\right)\nonumber\\
&+S_\text{m}\left\lbrack A^2(\phi) \tilde{g}_{\mu\nu}, \psi\right\rbrack\, . \label{EF_action}
\end{align}
Note that in this frame, the scalar field does not couple non-minimally to the curvature, but instead it appears in the matter action. The field equations found by varying the action (\ref{EF_action}) are:   
\begin{subequations}
	\begin{align}
	\tilde{R}_{\mu\nu}&=2\partial_\mu\phi\partial_\nu\phi-\frac{1}{2}\tilde{g}_{\mu\nu}U(\phi) \nonumber \\
	&+8\pi G\left(\tilde{T}_{\mu\nu} - \frac{1}{2}\tilde{T} \tilde{g}_{\mu\nu}\right)\, ,\label{EFieldEq1}\\
	\Box_{\tilde{g}}\phi&=-4\pi G\alpha(\phi)\tilde{T}+\frac{1}{4}\frac{\mathrm{d} U}{\mathrm{d}\phi}\, , \label{EFieldEq2}
	\end{align}
\end{subequations}
where the stress-energy tensor is given by $\tilde{T}^{\mu\nu}:=-2(-\tilde{g})^{-1/2}\delta S_\text{m}/\delta \tilde{g}_{\mu\nu}$\,. The function $\alpha(\varphi)=\partial \ln A(\varphi)/\partial \varphi$ gives a measure of the strength of the coupling between matter and the scalar field.

Note that the stress-energy tensor is no longer covariantly conserved, $\nabla_\mu\tilde{T}^{\mu\nu}\neq 0$\,. However, this does not mean that in this frame suddenly the WEP is no longer valid. To paraphrase \cite{ToGTs}, the WEP requires only that there is universal free-fall for test particles, not that the existence of this universality is immediately obvious by looking at the action. In particular, everything found to be physically valid in the Jordan frame must remain valid when we move to the Einstein frame \cite{STTofGrav}\,.

\subsection{Spontaneous scalarization}\label{SpontScal}

The prototypical screening mechanism is spontaneous scalarization \citep{SpontScal1, SpontScal2}\,. An intuitive understanding of the physics involved can be gained from the following approach, the basis of which is again in \cite{SpontScal1}.

First of all, to obtain GR solutions with a constant scalar field,  $\phi_0$ one can see, from equations (\ref{EFieldEq1}) and (\ref{EFieldEq2}), that $\alpha(\phi_0)=\left.\frac{\mathrm{d} U}{\mathrm{d}\phi}\right|_{\phi_0}=0$ must be imposed. Now consider an expansion of $\phi$ around this value to first order, $\phi=\phi_0+\epsilon \hat{\phi}$\,. Inserting this into the field equation (\ref{EFieldEq2}), we have:
\begin{equation}
\epsilon \Box \hat{\phi} = \epsilon \left\lbrack\frac{1}{4} \left. \frac{\mathrm{d}^2 U}{\mathrm{d} \phi ^2}\right|_{\phi_0} - 4\pi G T \left.\frac{\mathrm{d} \alpha}{\mathrm{d}\phi}\right|_{\phi_0}\right\rbrack \hat{\phi} \, .
\end{equation}
Thus the perturbation has a mass, $m_\phi^2=\frac{1}{4} \left. \frac{\mathrm{d}^2 U}{\mathrm{d} \phi ^2}\right|_{\phi_0}$. We also adopt the standard notation $\left.\frac{\mathrm{d} \alpha}{\mathrm{d}\phi}\right|_{\phi_0}=\beta$, and so,
\begin{equation}
\Box \hat{\phi}= (m_\phi^2-4\pi G T |\beta|)\hat{\phi} \left(+\mathcal{O}(\epsilon) \right):= -\mu^2 \hat{\phi} \, , \label{SpontScalEq1}
\end{equation} 
where $\mu$ is the effective mass of $\hat{\phi}$\,.

As a heuristic, consider a homogeneous ball of dust in flat space such that $4\pi G T|\beta|>m_\phi^2$\, then, after a Fourier transform, eq. (\ref{SpontScalEq1}) gives:
\begin{align}
\omega^2=\left(\frac{2\pi}{\lambda}\right)^2-\mu^2\, ,
\end{align}
where $\omega$ is the frequency of the variation of $\hat{\phi}$ with time, $\lambda$ is its wavelength and $\mu$ is real. Hence, for large enough $\lambda$, the frequency is imaginary and thus there is a \emph{tachyonic instability} of the scalar field. However, this instability will only be effective if the smallest unstable wavelength fits inside the matter distribution. 
When the stress energy tensor is proportional to the density, as is true for this heuristic and for low mass neutron stars, the critical value of $|\beta|$ for $m_\phi=0$ is determined by the stellar compactness $C$,
\begin{equation}
\left|\beta\right|\sim \frac{\pi R_\ast}{G M_\ast} =\frac{\pi}{C}\, .
\end{equation}
This is just a consequence of the facts that the threshold is for $\lambda = R_\ast$ and that $\mu^2 \propto \rho\propto M_\ast/R_\ast^3$ .

For static neutron stars with masses of about $1.4 \, M_\odot$\,, scalarization only occurs if $\beta \lesssim -4$ and is stronger the smaller $\beta$ is, as shown in \cite{SpontScal1}. When the scalar field grows, the stability is finally restored due to backreaction\,.

The most stringent bound stemming from binary pulsar experiments is that $\beta\gtrsim-4.3$ \cite{antoniadis2013massive}. Furthermore, as discussed e.g. in \cite{CosmoMLS}, to be consistent with cosmological observations one would need extreme fine-tuning to allow for massless scalars and spontaneous scalarization. 

One way out of these strong constraints is provided by adding a mass $m_\phi$ to the scalar field \citep{MassiveSpSc, Yazadjiev:2016pcb}\,. Within the limits $10^{-16}\,\si{eV}\ll m_\phi\lesssim 10^{-9}\,\si{eV}$\,, spontaneously scalarizing neutron stars would still be viable.

\subsection{Gravitational Higgs mechanism}

Using a toy model, it has been shown in \citep{GravHiggs1} and \citep{GravHiggs2} that spontaneous scalarization in neutron stars can lead to violations of the WEP in neutron star interiors. The model used to demonstrate this is a combination of a scalar-tensor Lagrangian in the Einstein frame and the Lagrangian of a \(U(1)\) field, $A_\mu$, which we shall take to be the electromagnetic field (in this section we use units defined by $c=G=\hbar=1$): 
\begin{equation}
\begin{split}
S_\text{GH}=&\frac{1}{4\pi}\int \mathrm{d}^4 x \, \sqrt{-g}\left\lbrace \frac{R}{4}-\frac{1}{2}g^{\mu\nu}\overline{D_\mu\phi}D_\nu\phi\right\rbrace\\
&-\frac{1}{4}\int \mathrm{d}^4 x \, \sqrt{-g} F_{\mu\nu}F^{\mu\nu}+S_\text{m}[A^2(\bar{\phi}\phi)g_{\mu\nu};A_\mu;\Psi^A]\, , \label{GravHiggsact}
\end{split}
\end{equation}
where $D_\mu\phi=\partial_\mu\phi-\mathrm{i} e A_\mu\phi$ is the gauge covariant derivative, $e$ is the charge, $F_{\mu\nu}=\nabla_{\mu}A_{\nu}-\nabla_{\nu}A_{\mu}$ is the usual field strength tensor and \(\Psi^A\) stands for all other matter fields. For compatibility with the gauge symmetry the conformal coupling depends only on the magnitude of $\phi$\,, i.e. $A=A(\bar{\phi}\phi)$ and we take $\exp{\left(\frac{1}{2}\beta \bar{\phi}\phi\right)}$ with constant $\beta$\,. The field equations obtained by varying the action (\ref{GravHiggsact}) are \citep{GravHiggs1, GravHiggs2}:

\begin{subequations}
	\label{FieldEquationsHiggs}
	\begin{align}
	&(\Box -e^2 A_\mu A^\mu-2\mathrm{i} e A^\mu\partial_\mu-\mathrm{i}e\nabla_\mu A^\mu)\phi=-4\pi T\beta\phi\, ,\label{FieldEquationsHiggs_a}\\
	&\nabla^\mu F_{\mu\nu}=J_\nu+J_\nu^{(\phi)}+m_\gamma^2(\bar{\phi}\phi)A_\nu\, ,\label{FieldEquationsHiggs_b}\\
	&G_{\mu\nu}=8\pi \left(T_{\mu\nu}+T_{\mu\nu}^{(\phi)}+T_{\mu\nu}^{(A)}+T_{\mu\nu}^{(\phi A)}\right)\, .\label{FieldEquationsHiggs_c}
	\end{align}
\end{subequations}

Where the following definitions have been used:

\begin{subequations}
	\begin{align}
	J_\mu=&-\frac{1}{\sqrt{-g}}\frac{\delta S_m}{\delta A^\mu}\, ,\\
	T_{\mu\nu}^{(\phi)}=&\frac{1}{4\pi}\left(\partial_{\mu}\bar{\phi}\partial_{\nu}\phi-\frac{1}{2}g_{\mu\nu}g^{\lambda\sigma}\partial_\lambda\bar{\phi}\partial_\sigma\phi  \right)\, ,\\
	J_{\mu}^{(\phi)}=&\frac{\mathrm{i}e}{8\pi}(\bar{\phi}\partial_\mu\phi-\phi\partial_\mu\bar{\phi})\, ,\\
	T_{\mu\nu}^{(A)}=&F_{\mu\lambda}F\indices{_\nu^{\lambda}} -\frac{1}{4}g_{\mu\nu}F_{\lambda\sigma}F^{\lambda\sigma}\nonumber\\
	&+m_\gamma^2({\bar\phi}\phi)\left(A_\mu A_\nu-\frac{1}{2}g_{\mu\nu}g^{\lambda\sigma}A_\lambda A_\sigma\right)\, ,\\
	T_{\mu\nu}^{(\phi A)}=&2\left(J_\mu^{(\phi)} A_\nu-\frac{1}{2}g_{\mu\nu}g^{\lambda\sigma}J_\lambda^{(\phi)}A_\sigma \right)\, .
	\end{align}
\end{subequations}

As stated the action (\ref{GravHiggsact}) is invariant under the \(U(1)\) transformation:
\begin{subequations}
	\begin{align}
	\phi&\rightarrow\phi\cdot\mathrm{e}^{\mathrm{i} e\lambda}\, ,\\
	A_\mu&\rightarrow A_\mu+\partial_\mu\lambda\, . \label{gaugeGH}
	\end{align}
\end{subequations}

Once a neutron star scalarizes, this symmetry is spontaneously broken which, through the Higgs mechanism, generates a photon mass,  $m_\gamma(\bar{\phi}\phi)\propto e^2 \bar{\phi}\phi$\,. That this necessarily leads to WEP violations is shown in Appendix \ref{TestWEPV}.

In \cite{GravHiggs2} it has been shown that for spherically symmetric, static cases, scalarized stars can be electrically neutral but still have a non-vanishing photon mass at the centre.
Even for a coupling charge as small as $e=\SI{e-36}{C}$\,, a photon mass of the order of several $\si{GeV}$ is generated. It is maximal in the centre of the scalarized star and approaches zero in the outer region. Thus, by introducing a coupling of the scalar field to the electromagnetic potential, a clear deviation from the standard model of particle physics arises in a neutron star's interior.

As we have mentioned before, within this model, the weak equivalence principle is violated. Through the action (\ref{GravHiggsact}), the scalar field does not only couple via the conformal factor in the Einstein frame. There does not exist a transformation which leads to a decoupling of $\phi$ from all the matter fields simultaneously. Hence, there is a fifth force.

\section{Impact of a gravitational Higgs mechanism on neutron star's magnetic fields}\label{NSModels}

In this section we study the impact of the scalar-Maxwell coupling on the magnetic fields of neutron stars. We will use the rather standard approximation of the magnetic field as a test field. This can be justified by comparing the energy contained by the test field to the other source fields and the gravitational binding energy of the non-magnetic background (see e.g. section 3 of \cite{Turolla:2015mwa}). It is also worth noting at this stage that, strictly speaking, we should also include the effect of the modified electrodynamics on the equation of state for the neutron star. Note however that, generically, including these effects would make the results less like GR and so only work towards the aim of this paper, to demonstrate that the effects are large, even for small couplings. 

Note that, as the universe is not static, the dynamics of the scalar field and the electromagnetic potential are also worth investigation.

\subsection{Approach \label{Approach}}

To compare the magnetic field of a neutron star in Einstein-Maxwell theory with that in the toy model, we first need to choose certain stellar properties which help us speak of comparable objects.

We use stars with equal ADM mass and equal values of the radial component of the magnetic field at the stellar radius, as these are two of the main observables of magnetized neutron stars. 

We assume that the matter inside of the neutron star behaves like a perfect fluid and thus the stress-energy tensor is given by:
\begin{equation}
T^{\mu \nu}_ \text{pf}=(\rho+P)u^\mu u^\nu+P g^{\mu\nu}\, ,
\end{equation}
with the pressure $P$, the density $\rho$ and the 4-velocity of fluid elements $u^\mu$\,.

The equation of state (EoS) that we use for both theories is:
\begin{equation}
P=K\cdot \rho^\Gamma \label{EoS}
\end{equation}
with $K=100$, $\Gamma=2$ and $\left[\rho\right]=\left[P\right]=\si{km^{-2}}$ (so we set $c=G=1$). This is a simple, classic model which still gives realistic solutions for neutron stars. 

For the function $A(\phi)$\,, which describes the coupling of the scalar field to matter, we use the quadratic model as detailed in section \ref{Theory}:
\begin{equation}
A(\phi)=\exp\left(\frac{1}{2} \beta \phi^2\right)\, .
\end{equation}
For generating comparable neutron star solutions, the first step is to choose a value for the central density of the neutron star in STT which is high enough to ensure scalarization. Next we determine a scalar field value, $\phi_\mathrm{c}$, at the stellar center which leads to a vanishing scalar field at infinity, $\phi_0 \rightarrow 0$\,. This is a choice we make to underline the possibility of large deviations inside the neutron star while still restoring GR far from the star.

We consider the Tolman-Oppenheimer-Volkoff (TOV) metric, i.e. the neutron stars we compute here are spherically symmetric and non-rotating:
\begin{equation}
\mathrm{d}s^2=-\mathrm{e}^{2 \Phi(r)} \mathrm{d}t^2+\mathrm{e}^{2 \Lambda(r)}\mathrm{d}r^2+r^2(\mathrm{d}\theta^2+\sin\theta\,^2 \mathrm{d}\varphi^2)\, .
\end{equation}
The standard equations we use to compute the behaviour of the scalar field $\phi(r)$, the TOV functions $\Phi(r)$ and $\Lambda(r)$ and the pressure $P(r)$ inside the neutron star can, for instance, be found in \cite{SpontScal1}.

With the relations $\nu(r)=2\Phi(r)$ and $\mu(r)=\frac{r}{2}\lbrack 1-\exp\lbrack-2\Lambda(r)\rbrack\rbrack$, $\phi_0$ can then be computed from values at the stellar surface (index s) \cite{SpontScal1}:
\begin{align}
\phi_0&=\phi_\text{s}+\frac{2\psi_\text{s}}{\sqrt{\nu_\text{s}^{\prime\, 2}+4\psi_\text{s}^2}}\,\mathrm{arctanh}\left[\frac{\sqrt{\nu_\text{s}^{\prime\, 2}+4\psi_\text{s}^2}}{\nu_\text{s}'+\frac{2}{R}}\right]\, .
\end{align}
One can also find the ADM mass $m_A$ from the surface values with:
\begin{align}
m_A=&\frac{R^2\nu_\text{s}'}{2}\sqrt{1-\frac{2\mu_\text{s}}{R}}\nonumber\\
&\exp{\left[-\frac{\nu_\text{s}'}{\sqrt{\nu_\text{s}'^2+4\psi_\text{s}^2}}\mathrm{arctanh}\left(\frac{\sqrt{\nu_\text{s}'^2+4\psi_\text{s}^2}}{\nu_\text{s}'+\frac{2}{R}}\right)\right]}\, .
\end{align}
When we have found a neutron star in scalar-tensor theory corresponding to a chosen central density $\rho_c$ with a certain ADM mass, we search for a star in GR which with a different value for $\rho_c$ reaches the same total mass. The gravitational field of the star is in this case computed by solving the TOV equations with $\phi(r)=\frac{\mathrm{d}\phi}{\mathrm{d} r}=0\;\forall\;r$.


Next, we study the magnetic fields of the computed stars. 
For this, we use the classical Maxwell theory in the GR case and consider a Proca-like theory where the photon mass varies with the scalar field for the scalarized star.
Following \cite{MagRelStars2}\,, we now introduce (in both theories) an electric current $J_\mu$ and the vector potential $A_\mu$ of the form:
\begin{align}
J_\mu=(0,0,0,J_\phi) \, , \quad  \mbox{and} \quad  A_\mu=(0,0,0,A_\phi)\, . \label{cursource}
\end{align}
Thus, we have a vanishing electric field, $E_\mu=0$\,, due to the assumption that ideal magnetohydrodynamics is valid here.

As in \cite{MagRelStars2}\,, we now assume that the current is given by a dipole, $J_\phi(r, \theta)=-j_1(r) \sin(\theta)^2$, and make the ansatz $A_\phi(r, \theta)=-a_1(r)\sin(\theta)^2$ for the potential.

For the external field of the star in Einstein-Maxwell theory there exists an analytic solution describing the behaviour of $a_1$ in the stellar exterior as a function of the magnetic dipole moment $\mu_\text{b}$ observed at infinity \citep{WasShap, MagRelStars2}:

\begin{align}
a_1^{(\text{ex})}=-\frac{3 \mu_\text{b}}{8 M^3}r^2\left[\ln \left(1-\frac{2M}{r}\right)+\frac{2M}{r}+\frac{2M^2}{r^2}\right]\, , \label{analytic}
\end{align}

As in \cite{MagRelStars3}\,, we calculate $a_1$\,, $a_1'$\,, $B_r$ and $B_\theta$ in units of the mean magnetic field strength, $\mu_b/R^3$\, .

In the interior of the neutron star in GR we have to solve the following differential equation (which is Maxwell's equation with the definitions for $A_\mu$ and $J_\mu$ inserted):
\begin{align}
-4\pi j_1(r)=&-\frac{2 a_1(r)}{r^2}+\mathrm{e}^{-2\Lambda(r)}a_1''(r)\nonumber\\
&+\mathrm{e}^{-2\Lambda(r)}a_1'(r) \lbrack\Phi'(r)-\Lambda'(r)\rbrack\, .
\end{align}
For the current function we use (see e.g. \cite{MagRelStars3}):
\begin{align}
j_1^{(\text{in})}=f_0 r^2(\rho+P)\, .
\end{align}
Here the constant $f_0$ has to be found by setting the values of $a_1(r=R)$ and $a_1'(r=R)$ equal to the ones we get from the analytic exterior solution (\ref{analytic}) and by requiring that the vector potential and its first derivative vanish at the stellar center.

As mentioned at the beginning of this section, our aim is to find equally massive neutron stars which have the same surface values for the radial component of the magnetic field both in STT and in GR. The relation between the vector potential and the radial component of the magnetic field at $\theta=0$ is \footnote{Note that, as in \cite{MagRelStars3}, we use here the tetrad components of the magnetic field.} \cite{MagRelStars3}:
\begin{align}
B_r=\frac{2 a_1}{r^2} \, .
\label{Br}
\end{align}

Hence, if we want to have $B_{r}^{\text{GR}}(r=R_\text{GR})=B_{r}^{ \text{STT}}(r=R_\text{STT})$, we have to set the vector potential at the boundary of the STT star to
\begin{align}
a_{1}^{\text{P}}(r=R_\text{STT})=a_{1}^{\text{MW}}(r=R_\text{GR})\cdot\frac{R_\text{STT}^2}{R_\text{GR}^2}\, , 
\end{align}
where the superscripts P and MW stand for Proca and Maxwell, respectively.

To find the derivative of $a_{1}^{\text{P}}$ at the stellar surface, it is necessary to know what asymptotic flatness would require of the magnetic field. Due to the mass arising from a scalar field which is not constant, one cannot simply use the standard results. Comparing the stress-energy tensor to the Einstein tensor one finds,
\begin{align}
m_\gamma^2(\bar{\phi}\phi) A_\mu A_\nu\propto \bar{\phi}\phi A_\mu A_\nu \sim \frac{1}{r^3}\, .
\end{align}
Going back to our neutron star model, we thus find $a_{1}^{\text{P}\,'}(r=R_\text{STT})$ by requiring that $a_1(r)\rightarrow 0$ \emph{at least} as fast as $1/\sqrt{r}$ for large $r$\,, because $\phi(r>R)\propto 1/r$ \cite{SpontScal1}. This also ensures that the stellar magnetic field approaches zero at high distances from the star.    

Now having determined the boundary values for our electromagnetic potential in the STT-Higgs case, all that is left to do is to determine the constant $f_0$ such that $a_1''$ vanishes at the stellar center and $a_1$ solves the Proca-like equations inside the star,
\begin{equation}
-4\pi j_1=\mathrm{e}^{-2\Lambda} \lbrack a_1''-a_1' (\Lambda'-\Phi')\rbrack-a_1\left(\frac{2}{r^2}+e_b\phi^2\right) \, .
\end{equation} 
Here, the constant $e_b$ is given by 
\begin{equation}
e_b=\left\lbrace\frac{\mu_0 c^4e^2}{4\pi G\hbar^2}\right\rbrace_\text{SI}=\left\lbrace\frac{e^2}{\hbar^2}\right\rbrace_\text{NGU}\, ,
\end{equation}
where the subscript NGU points to the system of \emph{natural geometric units} defined by $c=G=1$ and $\mu_0=4\pi$\,. 

The code with which the described steps are taken is written in Python. We use a fourth order Runge-Kutta method to solve the differential equations and a bisection method for root-finding.

\subsection{Results}\label{SecResults}
We have modelled the magnetic fields and electromagnetic potentials for three different setups:

\begin{enumerate}
	\item A neutron star in Einstein-Maxwell theory with the EoS given in equation (\ref{EoS})\,.
	\item A neutron star in STT without a photon mass with $\beta=-6$\,.
	\item A neutron star in STT with a gravitational Higgs mechanism, where $\beta=-6$ and the constant $e_b$ takes the three values 0.1, 1 and 10\,.  
\end{enumerate}

The point that we want to make,  that an alternative theory of gravity could have drastic consequences also on matter physics, is not strongly influenced by the precise choices of the scalar tensor theory. For our demonstrative purposes we thus find it justified to assume a massless $\phi$ and to use an already ruled out value for $\beta$. 

\begin{figure}[h!]
	\centering
	\includegraphics[width=0.48\textwidth]{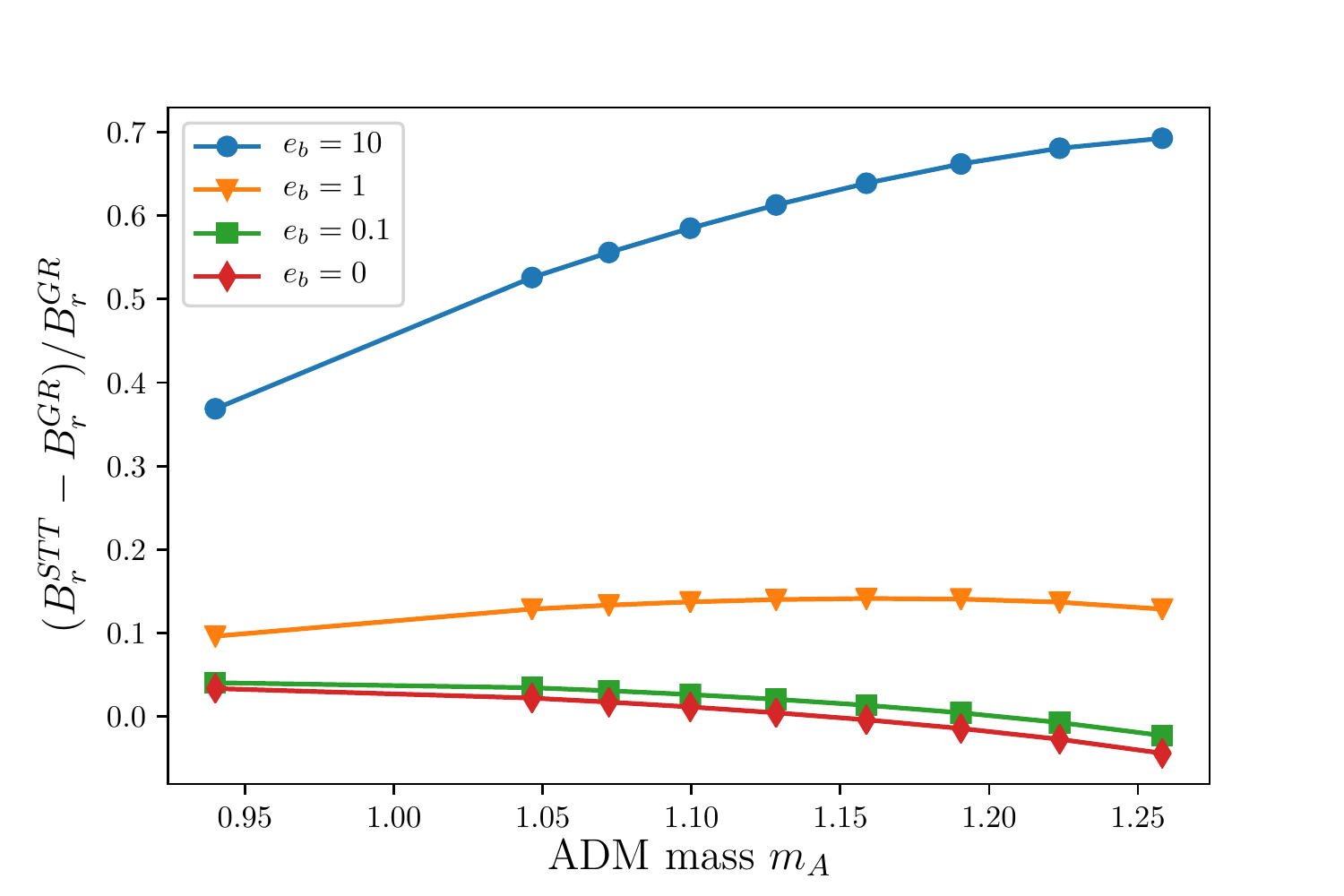}
	\caption{Ratio $(B_r^\text{STT}-B_r^\text{GR})/B_r^\text{GR}$ at $r=0.01 R$\label{ExFigC}}
\end{figure}
	
\begin{figure}[h!]
	\centering
	\includegraphics[width=0.48\textwidth]{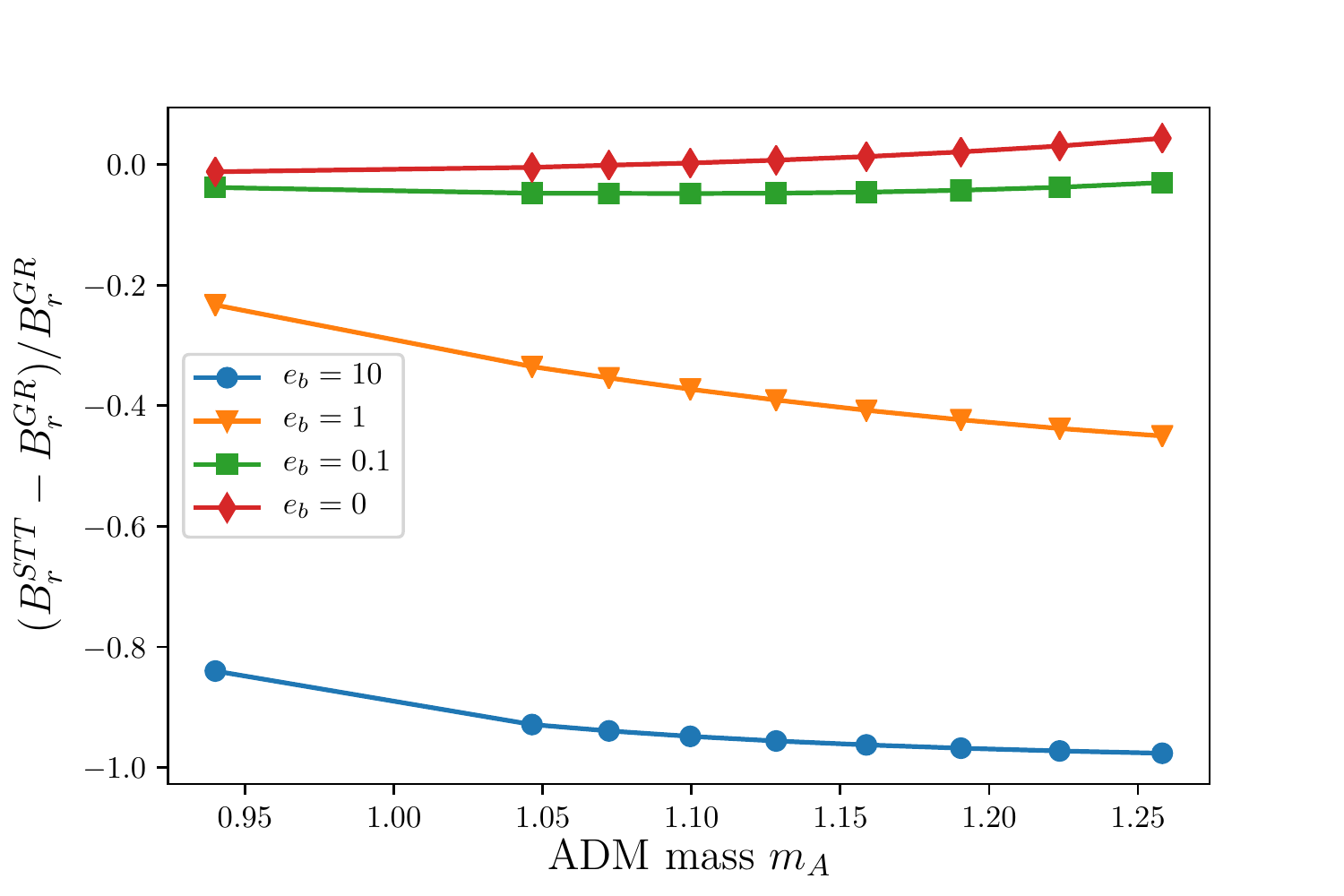}
	\caption{Ratio $(B_r^\text{STT}-B_r^\text{GR})/B_r^\text{GR}$ at $r=1.99 R$\label{ExFigD}}
\end{figure}

\begin{table}
	\centering
	\caption{Values used for $e_b$ and the corresponding values for $e$ and $m_\gamma/\phi$ in SI units and in units of electron charges $e_e$\,. \label{eVals}}
	\begin{tabular}{|c|c|c|c|c|}
		\toprule[1pt]
		$e_b$ & $e\,\lbrack\text{C}\rbrack$ & $e\,\lbrack e_e\rbrack$ & $m_\gamma/\phi\, \lbrack\text{kg}\rbrack$ &  $m_\gamma/\phi\, \lbrack\text{eV}\rbrack$ \\
		\midrule[1pt]
		0.1 & $9.59\times 10^{-54}$ & $5.98\times 10^{-35}$ & $1.11\times 10^{-43}$ & $6.23\times 10^{-8}$\\\addlinespace
		1.0 & $3.03\times 10^{-53}$ & $1.89\times 10^{-34}$ & $3.52\times 10^{-43}$ & $1.97\times 10^{-7}$\\\addlinespace
		10.0 & $9.59\times 10^{-53}$ & $5.98\times 10^{-34}$ & $1.11\times 10^{-42}$ & $6.23\times 10^{-7}$\\\addlinespace
		\bottomrule[1pt]
	\end{tabular}
\end{table}

Note that the photon mass is given by $m_\gamma\approx 3.5177\times 10^{-43} \sqrt{e_b} \phi\; \si{kg}$\, so, even for the highest value of $e_b$ plotted in Figs. \ref{ExFigC} and \ref{ExFigD}, the photon mass would be only of order $\SI{e-42}{kg}$, which is less than $\SI{1}{eV}$\,. Higher values have not been considered to avoid a loss of numerical precision. See Tab. \ref{eVals} for the charge and the photon mass corresponding to the three chosen $e_b$ \,.

\begin{figure}[h!]
	\centering
	\includegraphics[width=0.48\textwidth]{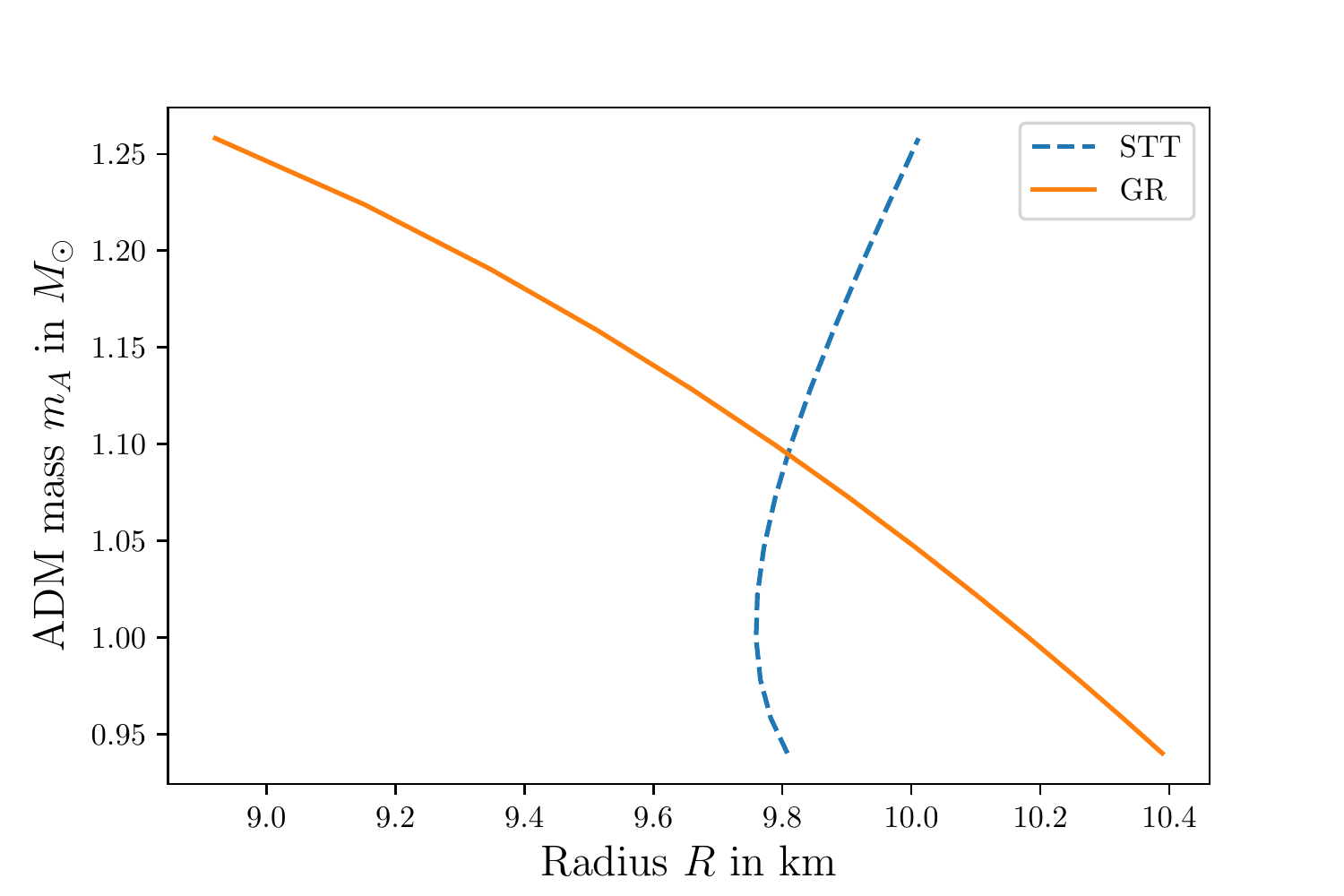}
	\caption{Mass and radius for a star in GR and in STT with the EoS given above. \label{MR}}
\end{figure}

As scalarized stars allow for higher masses and larger radii than those with the same equation of state in GR, we have only been able to compare the behaviour of the magnetic field for a certain range of values for $\phi$\,. The mass-radius curve for the stars considered here is shown in Fig. \ref{MR}\,. Additionally, we have listed the central values of the scalar field $\phi_c$ and the central densities $\rho_c$ in GR and STT corresponding to the various ADM masses in Table \ref{Prop}\,.

\begin{table}[h!]
	\centering
	\caption{Stellar properties for the ADM masses used here. \label{Prop}}
	\begin{tabular}{|c|c|c|c|}
		\toprule[1pt]
		$m_A$ $\lbrack M_\odot\rbrack$ & $\phi_c$ & $\rho_c^\text{STT}$ $\lbrack\si{kg\,km^{-3}}\rbrack$ & $\rho_c^\text{GR}$ $\lbrack\si{kg\,km^{-3}}\rbrack$\\
		\midrule[1pt]
		0.9400 & 0.1891 & $2.0\times 10^{27}$ & $1.33\times 10^{27}$\\\addlinespace
		1.0464 & 0.2335 & $2.5\times 10^{27}$ & $1.67\times 10^{27}$\\\addlinespace
		1.0723 & 0.2410 & $2.6\times 10^{27}$ & $1.77\times 10^{27}$\\\addlinespace
		1.0996 & 0.2481 & $2.7\times 10^{27}$ & $1.88\times 10^{27}$\\\addlinespace
		1.1284 & 0.2548 & $2.8\times 10^{27}$ & $2.01\times 10^{27}$\\\addlinespace
		1.1588 & 0.2610 & $2.9\times 10^{27}$ & $2.17\times 10^{27}$\\\addlinespace
		1.1905 & 0.2669 & $3.0\times 10^{27}$ & $2.36\times 10^{27}$\\\addlinespace
		1.2237 & 0.2723 & $3.1\times 10^{27}$ & $2.60\times 10^{27}$\\\addlinespace
		1.2582 & 0.2774 & $3.2\times 10^{27}$ & $2.92\times 10^{27}$\\\addlinespace
		\bottomrule[1pt]
	\end{tabular}
\end{table}

To evaluate the results of our code, we have plotted in Figs. \ref{ExFigC} and \ref{ExFigD} the deviations of the radial magnetic field component $B_r$ from the Einstein-Maxwell case for the various masses, divided by the respective value $B_r^\text{GR}$\,. All values are given in units of $\mu_b/ R^3$\,, where $R$ is the radius of the GR star.



\begin{figure}[h!]
	\centering
	\includegraphics[width=0.48\textwidth]{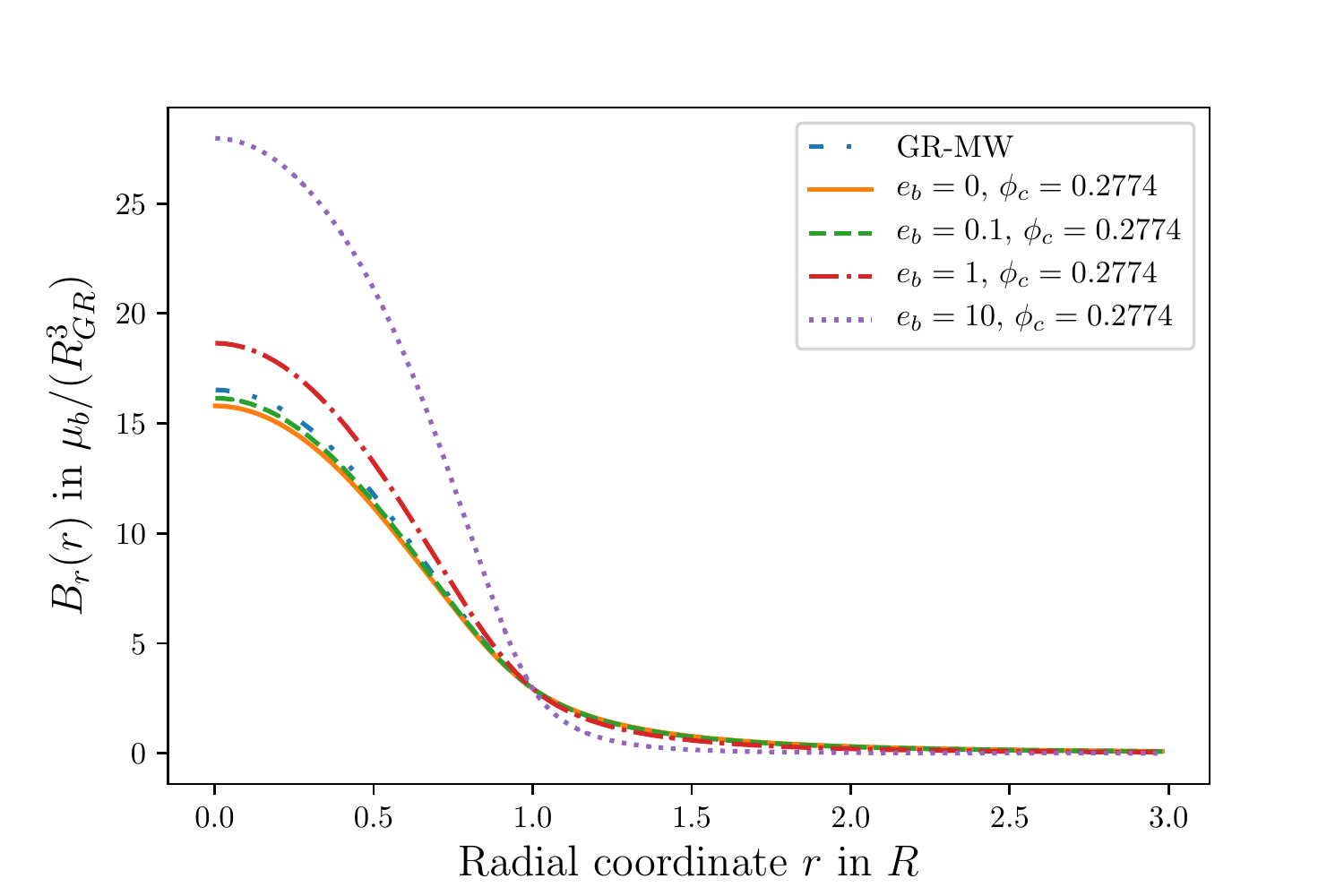}
	\caption{$B_r(r)$ for a neutron star with $m_A=1.2582\,M_\odot$\,. \label{BrCurve}}
\end{figure}

\begin{figure}[h!]
	\centering
	\includegraphics[width=0.48\textwidth]{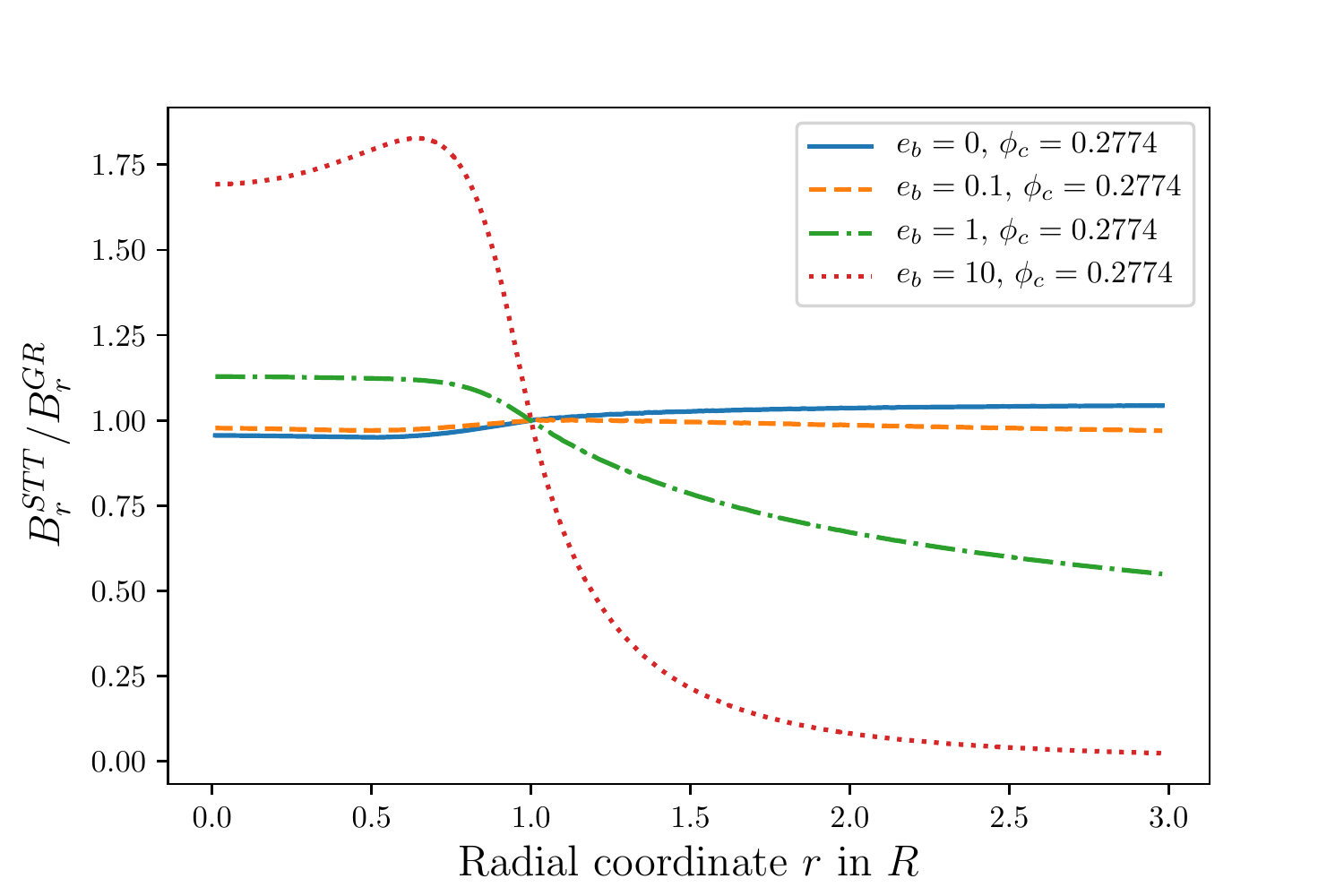}
	\caption{$B_r^\text{STT}/B_r^\text{GR}$ for a neutron star with $m_A=1.2582\,M_\odot$\,. \label{BrRat}}
\end{figure}

As can be seen in Fig. \ref{BrCurve}\,, for large $r\gg R$ the radial magnetic field approaches zero, as we have required it to. 




Also, the higher the photon mass, the more the radial magnetic field differs from the Einstein-Maxwell and the standard STT case in the vicinity of the star. This might be seen in a change of particle movement in the accretion disc and thus cause a variation in the emission spectrum of the neutron star.
When electrons move in a magnetic field, they are forced on a circular orbit and, due to the acceleration they feel, emit synchrotron radiation. 
The energy quanta emitted are
\begin{align}
E_\text{s}=\frac{\hbar e_e}{m_e c}B\, ,
\end{align} 
where $e_e$ is the electron charge, $m_e$ it's mass and $B$ the magnetic field in which it moves \cite{PulsarRad}.

So the ratio of energies in the case of two different magnetic fields, $B_a$ and $B_b$, is equal to the ratio of the field values, $\frac{E_a}{E_b}= \frac{B_a}{B_b}$\,. 

Thus, as in our models for stars with $e_b\neq 0$ we find that $(B_r^\text{STT}/B_r^\text{GR}) <1$ in the vicinity of the star, we expect a shift to lower frequencies in the synchrotron emission spectrum. 

\section{Discussion}\label{Discussion}

In this paper, we have compared the magnetic fields of neutron stars in Einstein-Maxwell theory and in STT with the gravitational Higgs mechanism. To do so, we chose neutron stars with equal ADM masses and equal surface values of the magnetic field's radial component. Our results show that a change of the photon's mass due to scalarization would indeed have a significant impact on magnetic fields.
Even for photon masses of the order of $10^{-42}\,\si{kg}$ we expect an increase of the central magnetic field by up to $70\%$\,. In the vicinity of the neutron star, the radial component of the magnetic field would be up to $80\%$ lower than what is expected from GR predictions. 

If the gravitational Higgs mechanism was in effect, we would thus expect a high discrepancy between the measured values of internal and external magnetic fields of neutron stars. The results gained using various observational methods would lead to seemingly contradictory assumptions on the magnetic field strength for the same star.


If predictions are made based solely on observations of the external field in the interior of, especially, a strongly magnetized neutron star (magnetar) there would be significant deviations which will affect  the interpretation of the quasi-periodic oscillation (QPO) spectra \citep{1998ApJ...498L..45D, 2007AdSpR..40.1446W} 
of both global magneto-elastic \citep{2011MNRAS.414.3014C, 2012MNRAS.423..811C, 2018MNRAS.476.4199G}
and/or localized crust oscillations \cite{2007MNRAS.375..261S} 
associated with the geometry and dynamics of the magnetic field.
During the last two decades, the modelling of the observed QPOs led to significant progress in associating the QPOs with the equation of state (EoS) and the strength and geometry of the magnetic field. These theoretical developments were based in classical general relativistic and sometime Newtonian estimations. A new magnetar hyper-flare \cite{Younes_2017} 
e.g. observed by NuSTAR hyper flare will be the ultimate test of the various approaches but also of the classical approach to the intensity and geometry of their interior magnetic field.

In the exterior of neutron stars the magnetic field is the main reason for being  observed and its speculated intensity and geometry is constraining their parameters. This means that, for example,  issues such as the dipole spin braking, the interpretation of the X-ray spectra and even the accretion models \citep{1997ApJS..113..367B,2019A&A...622A..61S} 
can be affected by the presence of the gravitational Higgs mechanism. Actually, a number of modern instruments are aiming in performing tests on the aforementioned potential issues \cite{2019SCPMA..6229506I}\,.
As an example, we refer to one of the main goals of NuSTAR related to accretion powered pulsars \citep{2016MNRAS.457..258T,2019A&A...622A..61S}. 
That is the  \emph{cyclotron resonance scattering} that is associated with quantum mechanical effects causing photons at specific energies to scatter away from the line of sight and in this way producing \emph{dips}  in the X-ray spectra. These energies are associated with the magnetic field strength and the knowledge of these values is directly linked with the associated accretion progress.

We thus emphasize that if the WEP was violated in the strong gravity regime, great care would be in order when interpreting observations of the interior or exterior magnetic fields of neutron stars. \\
As the model discussed here can only be considered as a toy model for demonstrative purposes, future research is needed to study in detail possible effects of alternative theories of gravitation beyond changing the laws of gravity. Furthermore, it is not within the realm of the present paper to propose concrete experiments and predict specific observations that would unambiguously point to the presence of a STT with a gravitational Higgs mechanism. 
By contrast, we aimed at providing a proof of principle revealing new promising directions of research in the field of alternative gravitational theories.

\acknowledgements{This work was supported by DFG research Grant No. 413873357. AC acknowledges financial support from the European Commision and T\"{U}B\.{I}TAK under the CO-FUNDED Brain Circulation Scheme 2.}
\appendix
\section{Demonstration of WEP violation in the Gravitational Higgs model \label{TestWEPV}}
Here we demonstrate that the motion of test particles in the Gravitational Higgs model is not universal. In particular, photons follow different paths than the rest of matter. To do so we first note that the rest of matter follows the geodesics on the, would be, Jordan metric,
\begin{equation}\label{GeoEqConf}
\tilde{u}^{\alpha}\tilde{\nabla}_{\alpha}\tilde{u}^{\beta}=0\,,
\end{equation}
where \(\tilde{\nabla}\) is the connection associated to the metric \(A^2(\bar{\phi}\phi)g_{\mu\nu}\). We use the \(U(1)\) gauge freedom to set \(\phi\) real and use the conformal factor \(A=\exp\left(\beta \phi^2/2\right)\), the basic conclusion is independent of these choices. Using the standard relations for conformal transformations we can rewrite equation \eqref{GeoEqConf} as,
\begin{equation}\label{GeoEq}
u^\alpha\nabla_\alpha u^\beta =- \beta \phi \left(\nabla_\alpha \phi\right) \left(g^{\alpha\beta}+ u^\alpha u^\beta\right)\,.
\end{equation}
We now compare this to the geometric optics limit of the equation for the \(U(1)\) field, following the notation of Section 22.5 of \cite{Misner:1974qy}. Note that naive application of this formalism for massive particles also takes the ultra-relativistic limit, as we are interested in the effects of the mass term one needs to be careful. To investigate the limit we are interested in we take,
\begin{equation}
A_\mu=\mathrm{Re}\left[\left(a_\mu + \epsilon b_\mu+\mathcal{O}(\epsilon^2)\right)\exp{\frac{i\theta}{\epsilon}}\right]\,,
\end{equation}
where \(\epsilon\) is a book-keeping parameter, and we must treat \(\phi\) as of order \(1/\epsilon\). Taking the divergence of equation \eqref{FieldEquationsHiggs_b} with the above ansatz we find, to lowest order,
\begin{equation}
a^\mu\nabla_\mu \theta\equiv a^\mu k_\mu =0\,,
\end{equation}
\textit{i.e.} in this gauge the photon has transverse polarization. Using this one can show that equation \eqref{FieldEquationsHiggs_b} reduces to,
\begin{equation}
k^2=-e^2 \epsilon^2 \phi^2\equiv -m^2_\gamma\,,
\end{equation}
So the wavevector is indeed timelike, as one would expect for a massive field. If we call the unit timelike vector in the same direction \(\ell\), then one can show (setting \(\epsilon \to 1\)),
\begin{equation}\label{photonpaths}
\ell^\alpha\nabla_\alpha \ell^\beta = - \left(\nabla_{\alpha} \log \phi\right)\left(g^{\alpha\beta}+\ell^\alpha\ell^\beta\right)\,
\end{equation}
which is not equivalent to the equation for other particles \eqref{GeoEq}\footnote{One can make them equivalent, by choosing the conformal factor $A= c \phi$, with $c$ a constant.}. Therefore the WEP is violated. As a final note, the logarithm should not cause concern for the behaviour of photons in the \(\phi\to 0\) limit of the theory as arriving at equation \eqref{photonpaths} requires assuming \(\phi\) is non-negligible and was calculated in the ``Proca gauge'', which is known to behave badly as \(m_\gamma\to 0\).

\section{Comparing radial magnetic fields of stars with equal compactness}

\begin{figure}[h!]
	\centering
	\includegraphics[width=0.48\textwidth]{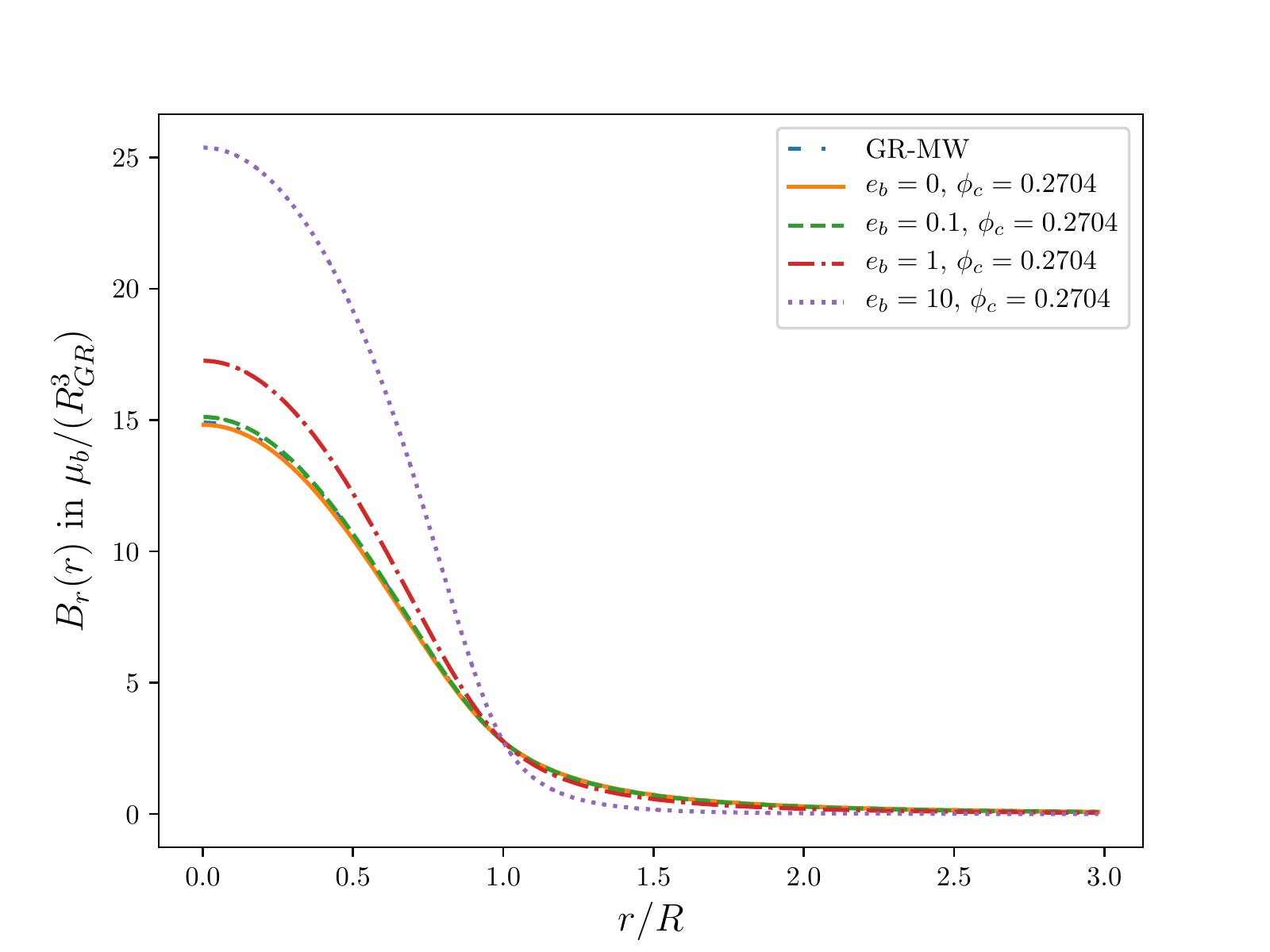}
	\caption{$B_r(r)$ for a neutron star with $m_A/R=0.18$\,. \label{BrCurve_eqComp}}
\end{figure}

\begin{figure}[h!]
	\centering
	\includegraphics[width=0.48\textwidth]{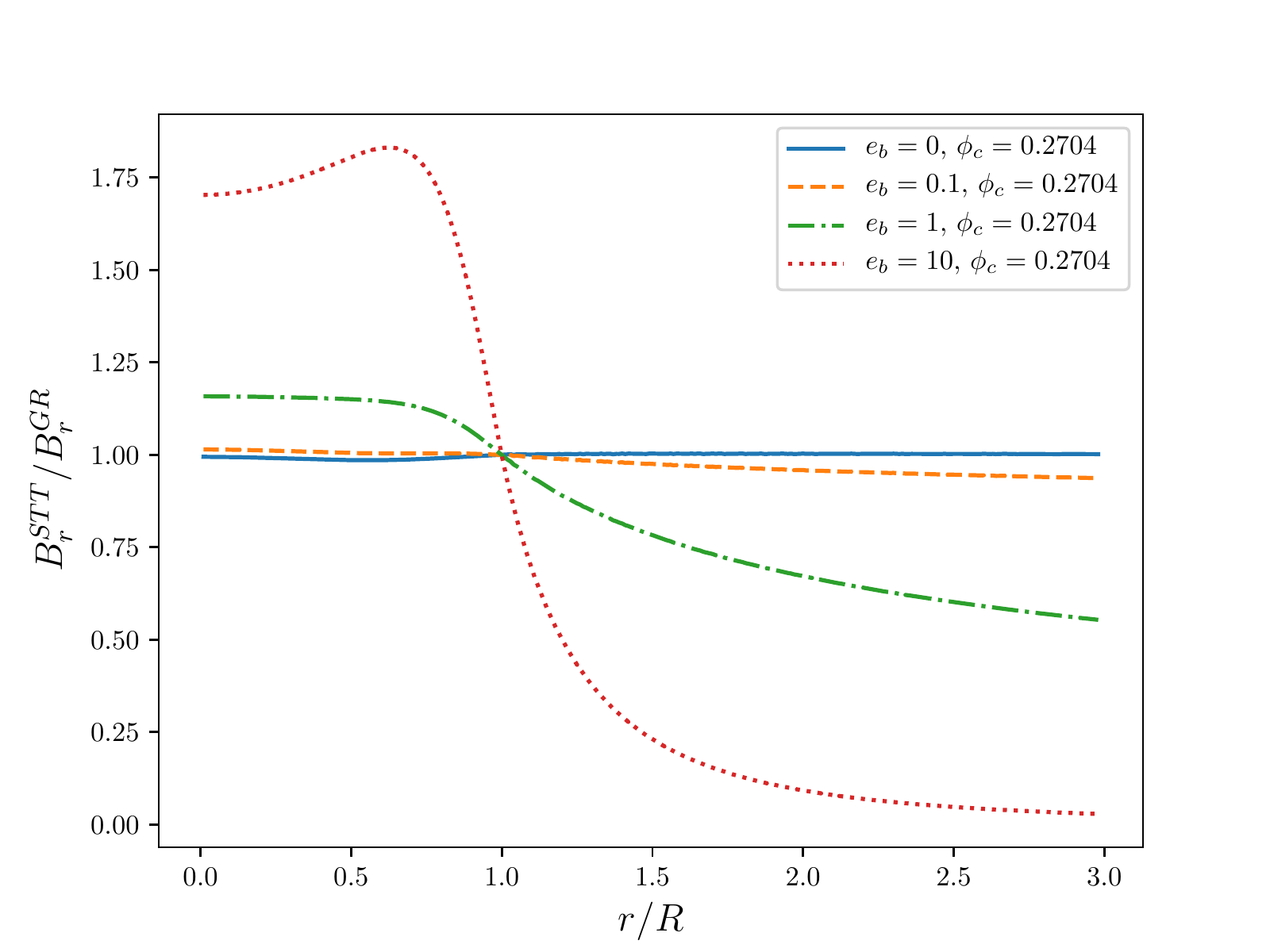}
	\caption{$B_r^\text{STT}/B_r^\text{GR}$ for a neutron star with $m_A/R=0.18$\,. \label{BrRat_eqComp}}
\end{figure}

In section \ref{SecResults} we have compared radial magnetic fields of neutron stars with equal ADM mass. A different plausible approach for observing the effect of STT with a gravitational Higgs mechanism on the magnetic field strengths would be to compare instead stars with equal compactness, i.e. $\left(M/R\right)^\text{GR}=\left(M/R\right)^\text{STT}$\,. To see whether a fixed compactness changes the main results presented in section \ref{SecResults}\,, we recreated the plots \ref{BrCurve} and \ref{BrRat} in figures \ref{BrCurve_eqComp}-\ref{BrRat_eqComp}\, for two stars with $\left(M/R\right)^\text{GR}=\left(M/R\right)^\text{STT}\approx 0.18$\,. As can be seen, the observations discussed in the main text of this paper stay valid also for this exemplary case, allowing us to assume that the chosen approach does not lead to unintended misinterpretations.   

~\\~\\

\newpage

\bibliography{literature}

\end{document}